\documentstyle[epsf,twocolumn,aps]{revtex}
\begin{document}
\twocolumn[\hsize\textwidth\columnwidth\hsize\csname @twocolumnfalse\endcsname
\title{Excitonic Correlations in the Intermetallic Fe$_2$VAl}
\author{Ruben Weht and W. E. Pickett}
\address{Department~of~Physics,~University~of~California,
Davis~CA~95616}
\date{\today}
\maketitle
\begin{abstract}
The intermetallic compound Fe$_2$VAl looks non-metallic in
transport and strongly metallic in thermodynamic and photoemission data.
It has in its band structure 
a highly differentiated set of valence and conduction bands leading
to a semimetallic system with a very low density of carriers.  
The pseudogap itself is due to interaction of Al states with the $d$
orbitals of Fe and V, but the resulting carriers have little Al 
character.  The effects
of generalized gradient corrections to the local density band structure
as well spin-orbit coupling are shown to be significant, reducing the 
carrier density by a factor of three.  Doping of this nonmagnetic compound
by 0.5 electrons per cell in a virtual crystal fashion results in a
moment of 0.5 $\mu_B$ and destroys the pseudogap.  
We assess the tendencies toward formation of an excitonic condensate 
and toward an excitonic Wigner crystal, and find both to be unlikely.
We propose a model is which the observed properties result from
excitonic correlations arising from two interpenetrating 
lattices of distinctive electrons
($e_g$ on V) and holes ($t_{2g}$ on Fe) of low density (one carrier of 
each sign per 350 formula units).
\end{abstract}

]
\section{Introduction}
The electronic behavior of Fe$_2$VAl has recently been discovered to
be highly unusual.\cite{nishino}  The resistivity 
increases by a factor of six
from 400 K to 2 K (where $\rho$=3 m$\Omega$ cm) characteristic of a 
non-metal, but other properties indicate metallic character.  A sharp
Fermi cutoff is observed in the photoemission spectrum, and the
specific heat coefficient $\gamma (T) \equiv$ C/T more than doubles
below 6 K (to 12 mJ/mol K$^2$).  Ferromagnetism (FM) in the related
compounds Fe$_3$Al and Fe$_3$Si and probable antiferromagnetism (AF) in
Fe$_2$VSi\cite{endo} suggests magnetic behavior  
may be responsible.  The stoichiometric compound itself has no magnetic
transition, although the alloy system Fe$_{2+x}$V$_{1-x}$Al and the
isoelectronic system Fe$_{2+x}$V$_{1-x}$Ga\cite{fevg1,fevg2} are
ferromagnetic with a Curie temperature that extrapolates to
zero as $x \rightarrow$ 0$^+$. 
The behavior of $\rho$ and C/T were suggested\cite{nishino}
as similar to those of heavy fermion metals, which would make it a candidate
for a $3d$ based heavy fermion system.  If the resistivity peaked and
began to decrease below 2K (as happens for UBe$_{13}$, for example) the
resemblance would be closer; however, the observed value of $\gamma$= 12
mJ/mole-K$^2$ at 2 K is well
below that of accepted heavy fermion materials.\cite{stewart} 

Although this compound falls within an alloy system where the V/Fe
ratio can be varied continuously, indications are\cite{nishino}
that the properties
noted above apply to the material near
or at the stoichiometric limit of an ideal Heusler (L2$_1$)
structure compound.  This structure type is
based on an underlying bcc lattice of lattice constant
$a/2$, with V at (0,0,0),
Al at ($\frac{1}{2},\frac{1}{2},\frac{1}{2})a$, and Fe atoms at
($\frac{1}{4},\frac{1}{4},\frac{1}{4})a$ and 
($\frac{3}{4},\frac{3}{4},\frac{3}{4})a$, where $a$=5.761~\AA~is
the lattice constant of the resulting fcc compound.
This structure is also that of Fe$_3$Al, which has two inequivalent Fe sites.
The Fe$_I$ site (which is the V site in Fe$_2$VAl) has 
eight Fe neighbors in octahedral configuration, while
while the Fe$_{II}$ site has four Fe and four Al neighbors.  
In Fe$_{2+x}$V$_{1-x}$Al the larger V atoms occupy the Fe$_I$ site
for $x\leq$0.  This trend is followed by all 
transition metal atoms to the left of
Fe in the periodic table (Ti, V, Cr).\cite{okpalugo}

Singh and Mazin 
\cite{djsiim} have reported local density functional results that
indicate that Fe$_2$VAl is a low carrier density,
compensated semimetal.  $\Gamma$-centered holes of Fe $t_{2g}$
antibonding character 
are compensated by zone-edge X-point holes
of V $e_g$ character.  Because the data cannot be interpreted simply
in terms of degenerate, noninteracting carriers with
this semimetallic band structure, Singh and Mazin suggest that the
behavior is caused by strong spin fluctuations of Fe atoms on the
V site, due to non-stoichiometry or to antisite defects.

In this paper we discuss other possible causes of the peculiar
observed behavior, starting from the semimetallic band structure obtained
in the local density approximation.  We show that Al plays a crucial,
but ultimately indirect, part in determining the electronic
structure, and that the magnetic state is unusually sensitive 
to band filling.  Based on a calculated carrier density
of one electron and one hole for each 350 unit cells, we consider parameters
related to a possible excitonic condensate or to an excitonic Wigner
crystal.  These exotic phases are
not favored by the parameters.
The character and density  of the carriers, the structure of
their sublattices, and the anticipated dielectric behavior suggest that
dynamic correlations of excitonic character are candidates to
account for the enhanced `metallic' spectral density that is accompanied
by decreasing conduction.

\section{Method of Calculation}
We have used the linearized augmented plane wave method\cite{djsbook}
that utilizes a fully general shape of density and potential.  Both
the WIEN97 code\cite{wien} and the WM-NRL code\cite{wmnrl}
have been used on different aspects
of the calculations.  The lattice constant of 5.761~\AA~was used. 
LAPW sphere radii (R) of 2.00 to 2.30 a.u. were used in various calculations,
and with very well converged basis sets no discernible effect of sphere
radius size was found.  Cutoffs of RK$_{max}$ up to 8.6-8.9 provided well
converged basis sets varying from 330 to more than
500 functions per primitive cell.  
Self-consistency was carried out on k-points meshes of around 200
points in the irreducible Brillouin zone
(12$ \times12\times 12$ and $15 \times 15 \times 15$
meshes). 

To assess the finer details of the predictions of density functional
theory, we used two forms of generalized gradient corrected 
exchange-correlation
functionals from Perdew and coworkers.\cite{gga}  
Gradient corrected functionals have been 
observed to give small but sometimes important corrections 
to the band structure in
several compounds.  Spin-orbit coupling was included in a second variation
approach as formulated by MacDonald {\it et al.}\cite{spinorb}

\section{Calculational Results}
\subsection{LDA Band Structure}
The band structure of Fe$_2$VAl is shown in Fig.~1. 
The crucial feature is a disjointedness between occupied valence
bands and unoccupied conduction bands that is unusual in an 
intermetallic compound. There are twelve bands to be filled by the
valence electrons.  Interestingly, these lower 
twelve bands are disconnected from
the remaining bands throughout the Brillouin zone.
The result is not quite a gap but rather 
a deep minimum (pseudogap) in the density of states (DOS) precisely where
the Fermi level (E$_F$) falls.  The density of states
N(E$_F$) does not vanish because there
is a very small overlap of a
V-derived conduction band minimum, containing electron carriers
at the three inequivalent X points, 
with a three-fold degenerate $\Gamma_{25^{\prime}}$
valence band maximum
arising from Fe-derived holes.  

The hole band at $\Gamma$ is roughly 40\% $t_{2g}$ on each Fe atom and 
15\% V, with minor Al character.
The electron pockets at X are V $e_g$ states (with minor Al character), 
but Fe $d$ character is
forbidden by symmetry.  The total and atom decomposed and projected DOS 
is shown in Fig.~2.  The Al states (not shown) 
are repelled from a wide region
including the Fermi level by the strong $d-d$ bonding, as discussed
by Singh and Mazin.  This repulsion, and the apparent unimportance
of the Al atom, leaves a network
that can be pictured as V-centered cubic VFe$_8$ clusters connected
along all edges in an fcc arrangement.  Contour plots of the holes and
the electrons in a (110) plane that contains the
nearest neighbor separations are shown in Fig.~3.

\begin{figure}[tbp]
\epsfxsize=9.0cm\centerline{\epsffile{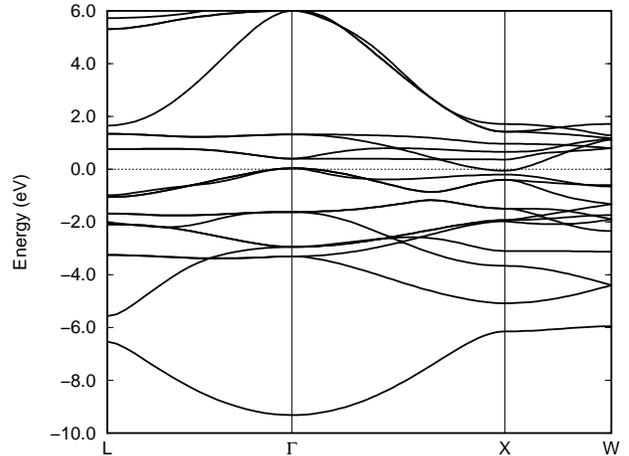}}
\caption{Full band structure of Fe$_2$VAl along principle symmetry
lines.  The dashed line denotes the Fermi level.
The lower band is $s$-like, primarily Al character.  The Fe and V
$d$ bands lie between -5 eV and 2 eV.
\label{Fig1}}
\end{figure}

\begin{figure}[tbp]
\epsfxsize=9.0cm\centerline{\epsffile{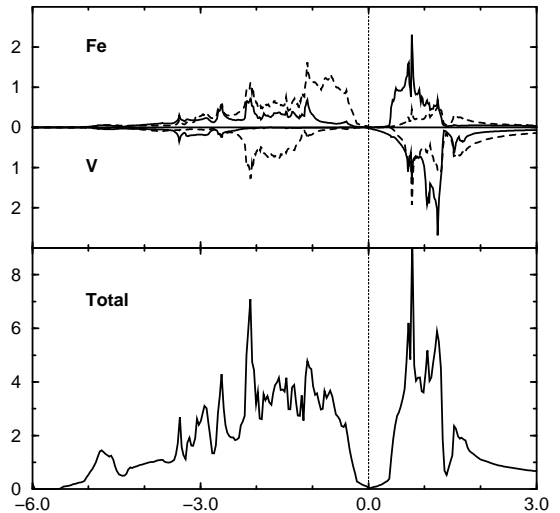}}
\caption{Density of states of Fe$_2$VAl (bottom), projected onto the
$e_g$ (solid lines) and $t_{2g}$ (dashed lines) crystal field
characters of the Fe and V atoms (top of figure).  The V density of
states are plotted downward.
Note that the Fermi level falls precisely in the minimum.
\label{Fig2}}
\end{figure}

\begin{figure}[tbp]
\epsfxsize=9.0cm\centerline{\epsffile{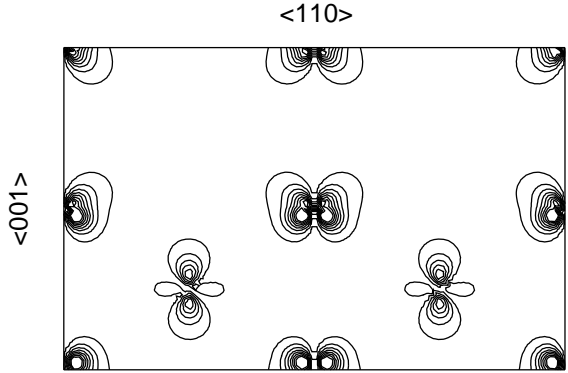}}
\caption{Contour plot of the Fe-centered hole charriers
(from the $\Gamma$ point
valence band maximum) and the $e-g$ type V-centered
electron carriers (from the X point minimum) in Fe$_2$VAl, plotted
in the (110) plane.  Al atoms are centered in the empty regions.
The separate densities were added to produce this figure.  The $t_{2g}$ type
density on Fe (corners, edges, and center of the plot) looks normal
when viewed in a (100) plane as is normally done.
\label{Fig3}}
\end{figure}

The DOS shown in Fig.~2 is projected onto $e_g$ and $t_{2g}$
states on both Fe and V.  The Fe $t_{2g}$ states lie almost
entirely below E$_F$, whereas the Fe $e_g$ DOS is split into roughly
equal amounts below and above the pseudogap.  For V the $e_g$ states lie 
entirely above the pseudogap, while the $t_{2g}$ states form 
bonding--antibonding complexes separated across the gap by 3 eV.

The minimum direct gap of $\approx$ 0.2 eV occurs along the (100)
directions near X.
A ``Penn gap'' between the occupied and
unoccupied DOSs, which gives a rough measure of where there is substantial
absorption weight, would be at least 2 eV.  The susceptibility 
of this system
is discussed below.

\subsection{Gradient Corrections and Spin-Orbit Coupling}
{\underline {Gradient Corrections}.}
The LDA band structure we obtain without gradient corrections and
neglecting spin-orbit coupling is indistinguishable from that of Singh
and Mazin.\cite{djsiim}
Since the fine details are on interest in determining the predicted
effective masses and carrier densities, we have calculated the
generalized gradient corrections of Perdew and coworkers\cite{gga} and
also included spin-orbit coupling.\cite{spinorb}  The resulting bands
are those shown in Fig.~1, and an enlargement
near the Fermi level is pictured in Fig.~4.  Gradient corrections
lead to a shift of the X point conduction minimum
upward with respect to the $\Gamma$ point band maximum, by 95 meV.
This reduces the band overlap (before spin-orbit coupling) from 200
meV to 100 meV, and reduces the number of carriers by roughly
(100/200)$^{3/2} \approx \frac{1}{3}$.

\begin{figure}[tbp]
\epsfxsize=9.0cm\centerline{\epsffile{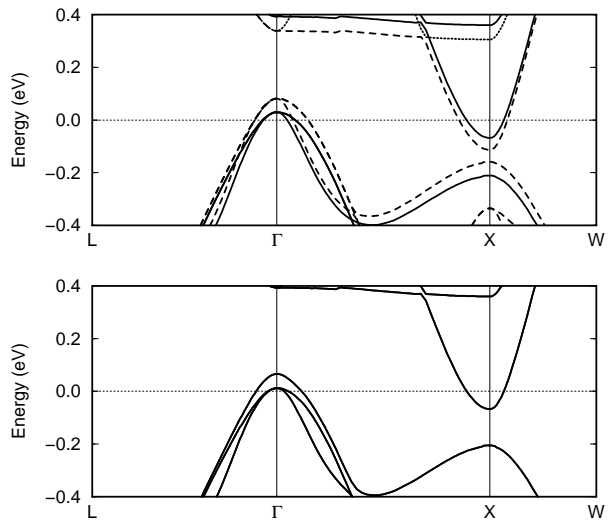}}
\caption{Enlargement of the band structure of Fe$_2$VAl near the Fermi
level.  Upper panel shows how gradient corrections affect the
band overlap (dashed lines show bands before gradient corrections
are included), by shifting states at X.  Both gradient
corrections and spin-orbit coupling are
included in the lower panel.
\label{Fig4}}
\end{figure}

{\underline {Spin-Orbit Coupling}.} The result of
spin-orbit coupling is to split the triply degenerate $\Gamma_{25^{\prime}}$
band maximum of mostly Fe $t_{2g}$ character
into $j=\frac{3}{2}$ and $j=\frac{5}{2}$ levels separated by 56 meV.  The
lower doublet leads to a pair of hole pockets whose occupation is
an order of magnitude less than that of the remaining hole pocket. 
The result is that the number of hole carriers is affected only little,
but the ``degeneracy" is lifted and the Fermi wavevector 
of the holes is increased
by 3$^{1/3}\approx$1.4.  

{\underline {Determination of Fermi Level}.} To determine
the position of the Fermi level, the hole pocket and the electron
ellipsoids at X were fit along high symmetry directions
to small wavevector expansions
\begin{equation}
\varepsilon_k = \varepsilon_o + \alpha k^2 + \beta k^4.
\end{equation}
Since the $k^4$ term is important even though the Fermi energy
is small, effective masses do not describe the dispersion precisely.
We obtain roughly
$m^*_h\approx1.0-1.1 m$ for the main hole pocket,
and $m^*_{\ell}=0.8 m$ and $m^*_{tr}=0.3-0.35
m$ for the longitudinal and transverse electron masses at X.
The band overlap is 130 meV, and compensation determines the Fermi
levels of 70 meV for the large hole pocket
and 60 meV for electrons (relative to their
respective band edges).  The DOSs of the various pockets are shown in
Fig.~5.

\begin{figure}[tpb]
\epsfxsize=8.0cm\centerline{\epsffile{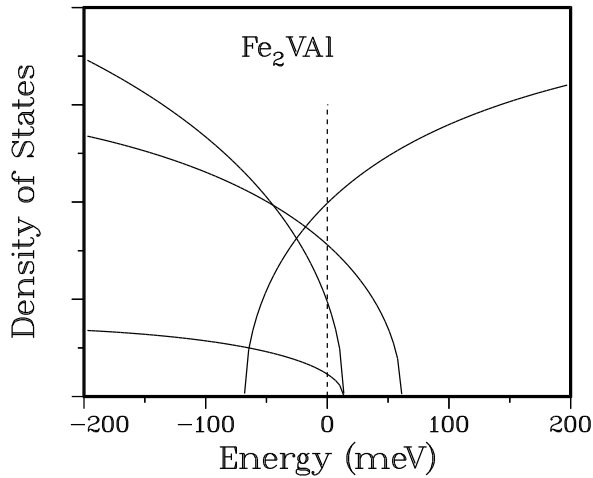}}
\caption{Relative densities of states of the overlapping bands very near
the Fermi level, from a small wavevector expansion discussed
in the text.  Two of the hole bands are nearly unoccupied and are
neglected in the discussion in the text.
\label{Fig5}}
\end{figure}

We finally obtain N(E$_F)\approx$0.1 states/eV, corresponding 
to a bare band specific heat coefficient $\gamma_b \approx$ 0.2 mJ/%
mole-K$^2$.  Compared to the extrapolated (T$\rightarrow$0) observed
value\cite{nishino} of 14 mJ/mole-K$^2$, 
the apparent enhancement factor of the
thermal mass is $m_{th}/m_b\approx$70.  We address this discrepancy
in Sec. IV.

{\underline {Mechanical Properties and Effect of Pressure.}
We calculated the energy for several crystal volumes.
The lattice constant that minimizes the energy is 0.7\% smaller than
the experimental value (V/V$_{obs}$=0.98).  A fit to a polynomial
gives a bulk modulus B=0.49 Mbar.  This indicates a relatively soft
lattice, which can be compared to 1.7 Mbar for Fe and 0.7 Mbar for Al.

It will be of interest for future experiments what the effect of 
pressure on the band overlap of the semimetallic state is.  From
calculations at V=1.025 V$_{obs}$ and V=0.95 V$_{obs}$ which translates
to a pressure difference of 36 Kbar, the change in band overlap 
is negligible.  Hence we predict that the effect of even substantial
pressure on the semimetallic state will be small.

\subsection{Magnetic Tendencies of Fe$_2$VAl}
We have checked for a
ferromagnetic instability by performing fixed spin moment calculations.
No such instability was found, consistent with the results of 
Singh and Mazin and
also with expectations based on the very small value of N(E$_F$), which
would leave the system far from a Stoner instability.
For small forced moments the V moment is parallel to that on the
Fe ion and the energy {\it vs.} moment curve is parabolic.  
For total moment greater than 0.3 $\mu_B$, however, the V
moment becomes antiparallel and saturates 
around 0.5 $\mu_B$ as the Fe moments
are driven to 2 $\mu_B$ and above.  This behavior is plotted in Fig.~6. 
In this range the energy {\it vs.} moment curve is linear.

\begin{figure}[tpb]
\epsfxsize=9.0cm\centerline{\epsffile{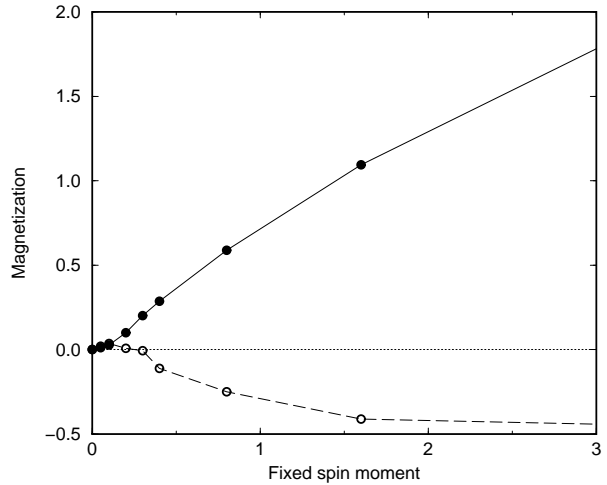}}
\caption{Behavior of the moments on the Fe atom (solid dots) and
on the V atom (open dots) when fixed spin moment calculations are
carried out.  Calculations were actually carried out
for moments above 3 $\mu_B$ as well as for the points indicated.
\label{Fig6}}
\end{figure}

A FM instability involves the $\vec q$ = 0 response of the
system, which is weak due to the very small Fermi surfaces.
The band structure contains more interesting low energy
response at the zone boundary (X point)
wavevector $\vec Q_{100}$=$(1,0,0)2\pi/a$, 
which could encourage an AF instability.  This response arises from 
the intersection of the Fermi surfaces when the hole-derived surfaces
at $\Gamma$ are displaced by $\vec Q_{100}$.
We searched for AF ordering of the simple
cubic Fe sublattice, with cubic (rocksalt) ordering of Fe moments. 
The energy increases
monotonically with Fe moment (arising from the imposed external
staggered field H$_{st}$) and no tendency toward instability was
found.  These results are consistent with the lack of any observed
magnetic order.

\subsection{Characteristics of the Semimetallic State}
The predicted carrier concentration is 2.9$\times$10$^{-3}$
carriers of each sign per formula unit (or one per 350 
$\approx~7 \times 7 \times 7$ primitive cells).
This number is four times smaller than that reported by Singh and Mazin due
to the effect of the generalized gradient corrections to LDA, which
push the electron band at X upward and thereby acts to decrease
the band overlap and hence the carrier concentration.
Since the electron carriers are V $e_g$ character, they reside on the V
sites which form an fcc lattice.  The hole carriers are of primarily
Fe $t_{2g}$
character, and the Fe sites form a sc array of lattice constant 
$\frac{a}{2}$, on which the holes are an average of (350)$^{1/3}$
$\frac{a}{2}$
$\approx 20$ \AA~apart.  

For qualitative purposes we can think in
terms of a mass $m_h^* \approx 1$ for holes with 1/700 of sites
occupied, and mass 
$m_e^* \approx 0.5$ of electrons with 1/350 of available sites occupied.
Taking the Drude plasma frequencies as $\Omega_p = 4\pi ne^2/m^*$ gives
$\hbar \Omega_p$=70 meV for the holes and 50 meV for the electrons.

\subsection{Importance of the Al Site}
Singh and Mazin\cite{djsiim} have noted that the Al $sp$ orbitals mix 
with the transition metal $d$ states, with the result that little Al
DOS remains in the vicinity of the Fermi level.  Since isoelectronic
Fe$_2$VGa has properties\cite{fevg1,fevg2}
similar to those of Fe$_2$VAl, we have carried
out parallel calculations on the Ga compound at its lattice constant
of 5.776~\AA.\cite{fevg1,fevg2}  The resulting band structure is very similar.  The
(primarily Ga) $s$ band centered 9.3 eV 
below E$_F$ is 1 eV lower than the Al band
in Fe$_2$VAl, but other differences are smaller.  The band overlap
giving rise to semimetallic character is 230 meV, compared to 130
meV in the Al compound.  As a result, Fermi surface sizes and carrier
concentrations are correspondingly larger than in Fe$_2$VAl.

To explore further the effect of the Al states, we calculated the
band structure of the fictitious compound Fe$_2$V in which 
the Al atom is simply removed.  Since this unit cell has an odd number
of electrons, the electronic structure must differ qualitatively; it
cannot be a paramagnetic semiconductor or be isomorphic to one 
(as the Fe$_2$VAl band structure is).  The resulting band structure is shown
in Fig.~7.  The difference in the
electronic structure is substantial, and is most obvious along the X-W
direction.  

\begin{figure}[tpb]
\epsfxsize=9.0cm\centerline{\epsffile{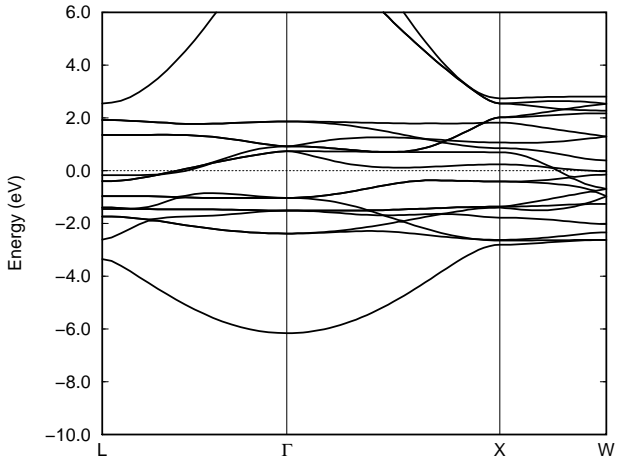}}
\caption{Band structure of the fictition compound Fe$_2$V, after
the removal of the Al atom.  The $d$ bands narrow by at least
1.5 eV, and the band topology near the Fermi level rearranges.
\label{Fig7}}
\end{figure}

The combined Fe-V $d$ bands are
considerably narrower, but corresponding bands can be
located throughout the Brillouin zone and throughout the valence/conduction
band energy range.  This congruity of the
band structures indicates
that removal of Al is analogous to removing three valence electrons
from the cell without removing bands (basis states available for
occupation).  As a result,
there is $d\rightarrow s$ charge promotion on both Fe and V as the
lowest, primarily $s$-like, band changes from largely Al character to
totally Fe and V character.
The filling of the $d$ bands drops, with an entire band along $\Gamma$-X-L
becoming unoccupied and two bands along $\Gamma$-L becoming half
empty.  Relative band shifts are also substantial, and resulting change
in band topology leads to the 
complete disappearance of the pseudogap due to band rearrangements at
the W point.  Thus the Al atom, or at least
its three electrons, play a crucial r\^ole in determining the electronic
behavior of Fe$_2$VAl.  The active states at the Fermi level 
nevertheless have only a small amount of Al character, as emphasized
by Singh and Mazin.\cite{djsiim}

\subsection{Doping on the V Sublattice}
As an indication of the effect of replacing V with a heavier
transition metal atom, a virtual crystal calculation was carried out
for the case where the charge on the V nucleus, and the number of 
electrons, are increased by 0.5.  This corresponds roughly to the case
of Fe$_2$V$_{1/2}$Cr$_{1/2}$Al, or more roughly to
Fe$_2$V$_{5/6}$Fe$_{1/6}$Al.  The result is a ferrimagnetic state
with a net moment near 0.5 $\mu_B$, {\it i.e.} almost equal to the
extra electronic charge in the cell.  The band structure, shown in Fig.~8 in
a region near the Fermi level, is severely disrupted near E$_F$, not resembling
at all a situation in which the bands are rigidly split in a Stoner
fashion.  The convergence to a self-consistent solution was very
slow, suggesting the FM state is not very stable.  
The net moment is
derived from 0.52 $\mu_B$ within each Fe sphere and $-0.45 \mu_B$ within
the V sphere.

We have looked at the same doping level by doing a virtual calculation
using the Al site, {\it i.e.} treating Fe$_2$VAl$_{1/2}$Si$_{1/2}$.
Considering the behavior we found in Sec. III.E, that Al seems to simply
dump its electrons into the system, this virtual crystal calculation
may be a very realistic treatment of the Al-Si alloy.  The result is 
very similar to doping on the V sublattice: the ferrimagnetic moment
is again near 0.5 $\mu_B$, with 0.42 on each Fe atom and -0.26 on each V atom.
The differences between doping on the V and Al sublattices can be
accounted almost entirely by the lowering of the V-derived electron
bands when doping is done on the V site.
We will report more of these results and a study of Fe$_2$VSi elsewhere.

\begin{figure}[tpb]
\epsfxsize=9.0cm\centerline{\epsffile{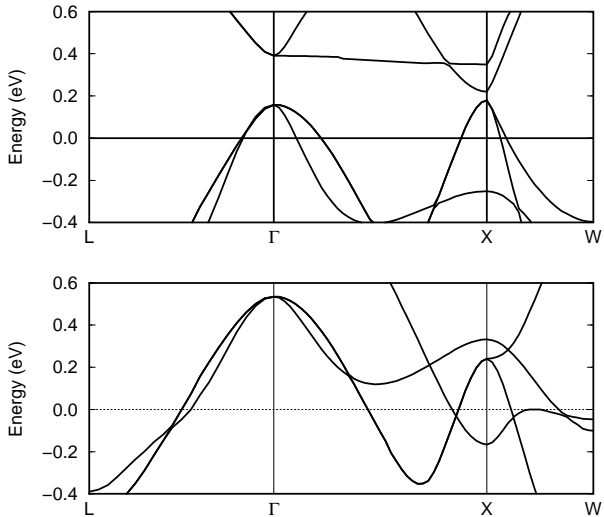}}
\caption{Majority (a) and minority (b) band structures of the
ferromagnetic state of the virtual crystal of Fe$_2$V$_{1/2}$%
Cr$_{1/2}$Al, illustrating the strong deviation from a
simple Stoner splitting of the bands in this system.
\label{Fig8}}
\end{figure}

\section{Discussion of Effects of Interactions}
\subsection{Excitonic Condensate?}
The study of semiconducting or semimetallic systems in which the
gap E$_g$ (positive or negative) is very small has a long history, with
residual interactions leading to the possibility of several exotic
phases.  A spontaneous condensed exciton 
phase is possible\cite{halprice} when the
gap E$_g$ is less than the 
the exciton binding energy E$_B^x$.   The 
occurrence of the spontaneous excitonic phase in a narrow gap 
semiconductor or low density semimetal has been extensively studied
and searched for experimentally, but only a few systems such as
TmSe$_{0.45}$Te$_{0.55}$ provide reasonable candidates for such a 
system.\cite{bucher}
Specific theoretical predictions are quite sensitive to the size and
shape of the electron and hole Fermi surfaces.\cite{kopaev}  
The possibility of an
exciton condensate in more highly correlated systems such as Kondo
insulators has been suggested by Duan, Arovas, and Sham.\cite{sham}

The classic case of the group V semimetals lies at one extreme,
where a one-electron picture works well and the effect of residual
interactions is small.  Bismuth is a compensated
semimetal with less than 10$^{-5}$ electron and hole carriers per atom,
yet it behaves at low temperature as a straightforward degenerate
metal with observable (but tiny) Fermi surfaces.   An intermediate
case is represented by ScN, a rocksalt structure semimetal with a 
LDA band structure very much like that of Fe$_2$VAl: a calculated
band overlap of 80 meV involving a threefold degenerate 
$\Gamma_{15}$ hole pocket
at $\Gamma$ and electron ellipsoids at X.  Monnier {\it et al.}%
\cite{monnier} considered the additional correlation energy 
of the electron and hole
carriers within LDA and concluded that 
ScN remains semimetallic ({\it i.e.} an electron-hole liquid) with
only modestly renormalized band overlap and effective masses.

The present case of Fe$_2$VAl has two new features not included in previous
models.\cite{halprice,kopaev}  First, 
the electrons and holes are well confined to
distinct interpenetrating lattices so that the 
discreteness of the lattice may have an effect.  Second, there
are likely to be strong residual interactions.  There will be repulsive
electron-electron and hole-hole interactions, not only on-site but
also intersite due to the weakness of the metallic screening.
On-site exchange interactions on Fe and V
may lead to magnetic fluctuations, although they should be much suppressed
due to the low value of N(E$_F$).
Then there will be the attractive electron-hole interactions.

The response is described by the dielectric function
$\varepsilon = 1 - \hat v \chi_o$ where $\hat v$ is a mean matrix element
of the Coulomb interaction
and the non-interacting
susceptibility is
\begin{equation}
\chi_o(\vec q,\omega)=2 \sum_{k,n,m} \frac{f(\varepsilon_{k,n})
-f(\varepsilon_{k+q,m})}{\varepsilon_{k+q,m}-\varepsilon_{k,n}
-\hbar \omega},
\end{equation}
where $\varepsilon_{k,n}$ is the band energy of band $n$ and $f(\varepsilon)$
is the Fermi-Dirac thermal occupation factor.
For the band structure of Fe$_2$VAl there are three type of contributions
at low energy $\hbar \omega$.  For $\vec q \rightarrow 0$ 
there is the usual repulsive 
intraband scattering (from holes and electrons separately)
that gives rise to Drude absorption in the
optical spectrum.  This response corresponds to a Lindhard-like
susceptibility from both the hole and electron pockets, with Drude
plasma energies given in Sec. III.D.  The characteristic Fermi
wavelengths are of the order of $\lambda_F \approx 12a$=70~\AA.

Second, near the X point where 
$\vec q$ = $\vec Q_{100}$ 
and equivalent points there will be low $\omega$ response as the
$\Gamma$-centered hole Fermi surface displaced by $\vec Q_{100}$ intersects
one of
the electron ellipsoid Fermi surfaces.  Third, there will be low energy
intervalley scattering of electrons centered at $\vec Q_{110}$ = 
(1,0,0)$\frac{2\pi}{a}$ - (0,-1,0)$\frac{2\pi}{a}$ and symmetry-related
vectors.  In the fcc lattice $\vec Q_{110}$ reduces to $\vec Q_{100}$. 
Thus attractive electron-hole scattering and intervalley 
repulsive electron-electron scattering
occur at the same reduced wavevectors.

\subsection{Excitonic Parameters}
The scale factors describing the excitonic system\cite{monnier}
depend on the background
dielectric constant $\epsilon_{\infty}$ for the rigid lattice in the
absence of carriers, and on the reduced mass $\mu$ given by
\begin{equation}
\frac{1}{\mu}=\frac{1}{m_e^*} + \frac{1}{m_h^*},
\end{equation}
where the electron effective mass $m_e^*$ is given by
\begin{equation}
\frac{3}{m_e^*} = \frac{1}{m_{\ell}} + \frac{2}{m_{tr}}
\end{equation}
in terms of the longitudinal ($m_{\ell}$) and transverse ($m_{tr}$)
masses of the electron ellipsoid.  For the calculated bands we
obtain $\mu\approx\frac{1}{3}m$.  The density of electron-hole pairs
corresponds to a (bare, see below) density parameter $r_s$=30.
The dielectric constant is difficult to estimate.  It involves
dipole matrix elements between $d$ bands, which vanish in the atomic
limit (dipole transitions require $p\rightarrow d$ or
$d \rightarrow f$ transitions in this limit).  Moreover, the valence and
conduction bands are dominated by Fe and V respectively, 
on two separate sublattices, which will tend to reduce matrix elements.
Since the Fe and V $d$ states retain
much of their atomic character in this solid, we do not expect
a large value of $\epsilon_{\infty}$.  

The exciton radius $a_B^x$, effective Rydberg E$_B^x$ (the binding
energy), and density parameter $r_s^x$ are given by\cite{monnier}
\begin{equation}
a_B^x = \epsilon_{\infty} \frac{m}{\mu} a_B,
\end{equation}
\begin{equation}
E_B^x = \frac{1}{\epsilon_{\infty}^2} \frac{\mu}{m} E_R,
\end{equation}
and
\begin{equation}
r_s^x = \frac{a_B}{a_B^x} r_s = \frac{1}{\epsilon_{\infty}}
\frac{\mu}{m} r_s,
\end{equation}
where E$_R$ is the Rydberg and $a_B$ is the Bohr radius.

In Table I we provide values of these parameters for $\epsilon_{\infty}$
=2, 5, 10, and 20.  From the band structure and the band character we
expect $\epsilon_{\infty}$ to lie in the lower end of this range.
In any case, however, the exciton radius is large and
the resulting effective density parameter $r_s^x$
is far from the low density range where an exciton condensate might
be expected.  Moreover, the Thomas-Fermi screening length 
for this density of electrons (or holes) is still
$\approx$2~\AA, so screening 
is still a consideration.  In any case, a true exciton condensate would be
an insulator, which is not the case for the reported samples (unless the
conduction is a result of defects). 
Thus an exciton condensate is not expected to arise.  Moreover, since there
is only a factor of two difference in masses, a hole Wigner crystal screened 
by the lighter electrons also is not a likely scenario.  However,
an electron-hole plasma of low
density ($r_s$=30) may have strong dynamic correlations, which we address next. 

\begin{table}
\caption[]{
Values of the excitonic radius $a_B^x$, binding energy $E_B^x$,
and density parameter $r_x^x$ for representative values of
$\epsilon_{\infty}$.
}
\begin{tabular}{cccc}
$\epsilon_{\infty}$ & $a_B^x/a_B$ & $E_B^x$ (meV) & $r_s^x$ \\
\tableline
   2    &   6  & 1150 & 5  \\
   5    &  15  &  180 & 2  \\
  10    &  30  &  45 &  1 \\
  20    &  60  &  11 & 0.5 \\
\tableline
\end{tabular}
\label{tableI}
\end{table}

\subsection{Excitonic Correlations}
An interaction that will be nearly unscreened in this 
system is the attraction
between an electron on a V site and a hole on a neighboring Fe site.
The V-Fe separation is only 2.5~\AA, and since the density of electrons
and holes separately is in the vicinity (if $\epsilon_{\infty}$
is small) of that where Wigner crystallization
should occur,\cite{wigner} a homogeneous screening
approximation may have broken down anyway.  We suggest that a 
Hamiltonian of the form
\begin{eqnarray}
H & = & \sum_{i,i^{\prime},s,s^{\prime}} t^e_{i,i^{\prime}} 
                e^{\dag}_{i,s} e_{{i,s}^{\prime}} +
  \sum_{j,j^{\prime},s,s^{\prime}} t^h_{j,j^{\prime}} 
                h^{\dag}_{j,s} h_{{j,s}^{\prime}} \\ \nonumber
 & & -V^{eh}\sum_{<i,j>,s,s^{\prime}}n^e_{i,s} n^h_{j,s^{\prime}} \\ \nonumber
 & & +U^e \sum_i 
          n^e_{i,+} n^e_{i,-}
 +U^h \sum_j
          n^h_{j,+} n^h_{j,-}
\end{eqnarray}
includes the important residual interactions.  Here $e^{\dag}_{i,s}$
creates an electron of spin $s$ (equal to + or -)
on the $i$-th site of the V sublattice, and
$h^{\dag}_{j,s}$ creates a hole of spin $s$
on the $j$-th site of the Fe sublattice, $n^e=e^{\dag}e$ and similarly for
the hole occupation operator.
The constants $V^{eh}$ and $U^e, U^h$ represent the corresponding interactions
confined here to on-site for like particles, or nearest neighbors for
the electron-hole interaction.
The hopping amplitudes $t^e$ and $t^h$ are determined so as to reproduce
the dispersion seen in Fig.~4.  
The attractive $e-h$ attraction $V^{eh}$ might be of the order of
\begin{eqnarray}
V^{eh}\geq \frac{e^2}{\epsilon_{\infty} r_{n.n.}} e^{-k_{TF}r_{n.n.}} 
      = \frac{2.5 eV}{\epsilon_{\infty}} \times 0.3 
       \approx  150~meV 
\end{eqnarray}
if $\epsilon_{\infty} \approx 5$, with the equality being
applicable only if Thomas-Fermi screening 
is applicable.  If both dielectric and Thomas-Fermi screening are
inapplicable due to the short separation, $V^{eh}$ could approach the
1-2 eV range.  

Falicov and Kimball\cite{falicov} considered such a
model, but treated only the case of infinitely heavy holes and neglect
of identical particle repulsion.  They obtained an anomalous temperature 
dependence of the carrier concentration.  Kasuya\cite{kasuya} has considered 
the possibility of Wigner crystallization of excitons in a related picture,
and suggested that picture might be appropriate for rare earth pnictides
such as LaSb.  The situation presented by Fe$_2$VAl does not seem to
fit well within either of these pictures.

Since we expect the kinetic energy to be sufficient to inhibit formation
of bound excitons (as discussed in Sec. IV.B), 
dynamic correlations of excitonic character may
be substantial.  To account for the observed behavior, the interactions
have to be large enough to keep the system from simply reverting to
a degenerate metal (such as Bi).  Two effects oppose formation of coherent propagating
electron-hole pairs: the electrons and holes have different velocities
(although very similar in magnitude),
and they move on inequivalent sublattices.  Nevertheless excitonic
correlations should occur and grow in strength
as the temperature is lowered.  These correlations
result in neutral objects that do not carry current, hence the resistivity
should increase at low temperature, as observed, as carriers are
effectively removed in pairs from the conduction process. 
Whether the photoemission spectra and the thermal mass enhancement
can be obtained will require further treatment of this model Hamiltonian,
which will
be presented elsewhere.  This Hamiltonian would be expected to show
superconductivity in its phase diagram as well, although perhaps not at
very low density.

\section{Summary}
Within the local density approximation, Fe$_2$VAl is found to be
a low density semimetal.  It is not unstable toward magnetic ordering,
in agreement with observation, although addition of rather small amounts of
carriers produces ferromagnetism, as observed.  The Al atom is found
to have a strong but indirect effect on the electronic structure
of this compound.  The observed properties, including a heat
capacity that is seventy times larger than the semimetallic DOS
can account for, indicate that other processes are occurring in
this system.  

The density of electrons and holes is in the range where individually
formation of a Wigner crystal is expected, and hence long-range
interactions might be expected to be of importance.
We have suggested rather that dynamic exciton-like short-range correlations
(arising from short-range interactions)
will dominate the observed behavior, and account for at least some
of the enhancement of the thermal mass and for the upturn
in the resistivity at low temperature.  We have considered the
possibility of an exciton condensate ground state, but for this 
picture to be viable,
but even for an unreasonably small background dielectric constant  
a conventional condensate does not seem to be indicated.  
A two-band, extended Hubbard model 
with attractive electron-hole interaction has been suggested
as a possible model to capture the essential processes, including
excitonic correlations.  The possibility of inhomogeneous magnetism, as
suggested by Singh and Mazin, remains a possibility for accounting
for some of the observed behavior.

\section{Acknowledgments}
We are grateful to W. L. Lambrecht for bringing Ref.\cite{monnier} to 
our attention.
This work was supported by the U.~S.~Office of Naval Research.
 

\end{document}